\documentclass[11pt]{article}
\usepackage{a4wide,epsfig,amsmath,amssymb,cite,scalefnt}

\newcommand{\eps}{$\varepsilon$}
\newcommand{\epscalar}{$\varepsilon$-scalar}

\newcommand{\abbrev}{\scalefont{.9}}
\newcommand{\drbar}{$\overline{\mbox{\abbrev DR}}$}
\newcommand{\msbar}{$\overline{\mbox{\abbrev MS}}$}

\newcommand{\betaMSbar}{\beta^{\overline{\rm MS}}}

\newcommand{\apiDR}{\frac{\asDRbar}{4 \pi}}
\newcommand{\apiMS}{\frac{\asMSbar}{4 \pi}}

\newcommand{\asDRbar}{\alpha_s^{\overline{\rm DR}}}
\newcommand{\asMSbar}{\alpha_s^{\overline{\rm MS}}}

\newcommand{\ayMSbar}{\hat{\alpha}_s^{\overline{\rm MS}}}
\newcommand{\gsMSbar}{g_s^{\overline{\rm MS}}}
\newcommand{\gsDRbar}{g_s^{\overline{\rm DR}}}

\newcommand{\ala}{\frac{\alpha_L^{A}}{4\pi}}
\newcommand{\alra}{\frac{\alpha_{LR}^{A}}{4\pi}}
\newcommand{\alab}{\frac{\alpha_L^{AB}}{4\pi}}
\newcommand{\alrab}{\frac{\alpha_{LR}^{AB}}{4\pi}}

\newcommand{\dreg}{{\abbrev DREG}}
\newcommand{\dred}{{\abbrev DRED}}

\newcommand{\mgDRbar}{m_{\tilde g}^{\overline{\rm DR}}}
\newcommand{\mgMSbar}{m_{\tilde g}^{\overline{\rm MS}}}
\newcommand{\mqDRbar}{m_{q}^{\overline{\rm DR}}}
\newcommand{\mqMSbar}{m_{q}^{\overline{\rm MS}}}
\newcommand{\cA}{C_A}
\newcommand{\cR}{C_F}
\newcommand{\cF}{C_F}

\begin{document}

\title{\vskip-3cm{\baselineskip14pt
    \begin{flushleft}
      \normalsize SFB/CPP-09-75 \\
      \normalsize TTP09-29
  \end{flushleft}}
  \vskip1.5cm
  Two-loop parameter relations between dimensional regularization and
  dimensional reduction applied to SUSY-QCD
}
\author{
  L. Mihaila$$\\[1em]
  {\small\it  Institut f{\"u}r Theoretische Teilchenphysik,
    Universit{\"a}t Karlsruhe,}\\
  {\small\it Karlsruhe Institute of Technology (KIT)}\\
}

\date{}

\maketitle

\begin{abstract}
  The two-loop relations between the running gluino-quark-squark
  coupling, the gluino and the quark mass defined in dimensional
  regularization (DREG) and  dimensional reduction (DRED) in the
  framework of SUSY-QCD are presented. Furthermore, we verify with the
  help of these relations that the three-loop $\beta$-functions derived
  in the minimal subtraction scheme combined with DREG or DRED transform into
  each other. This result confirms the equivalence of the two schemes at the
  three-loop order, if  applied to SUSY-QCD. 
\medskip


\end{abstract}

\section{\label{sec::intro}Introduction}

\dred{} has been introduced in Ref.~\cite{siegel}
as a regularization scheme for supersymmetric gauge theories which
maintains supersymmetry (SUSY) and at the same time retains the elegant
features of \dreg{}\cite{'tHooft:1972fi}, especially the gauge
invariance. The essential difference between \dred{} and \dreg{}
is that the continuation from $4$ to $D$ dimensions  is 
made by {\it compactification\/}. After {\it dimensional reduction} to
$D=4- 2\varepsilon$, it is only the $D$ components of the gauge field that
generate the actual gauge interactions. The remaining $2\varepsilon$ components
behave under gauge transformations as a multiplet of scalar fields, usually
 called  \epscalar s. 

As pointed out by Siegel himself~\cite{siegelb}, there are potential
problems with \dred{}. In Ref.~\cite{Avdeev:1981vf} it has been shown that
the variation $\delta S$ of the action of a pure (no chiral matter)
supersymmetric gauge theory is nonzero even with \dred.
If $\delta S$ gives a nonzero result when inserted in a Green's function this
creates an apparent violation of supersymmetric Ward identities.
Within \dreg{} this happens at one-loop order. On the other hand, within
\dred{} all explicit 
calculations up to two-loop order have found zero for such 
insertions~\cite{jjs,ds}. Recently,
 a mathematically consistent formulation of \dred{}~\cite{ds} and 
rigorous methods  to prove its supersymmetric properties
~\cite{Hollik:2005nn} have been introduced.
\\
Another way to verify the consistency of \dred{} with SUSY
is to study the behaviour under the renormalisation of the
\epscalar-couplings (also called {\it evanescent} couplings) to
matter and gauge fields. In a
supersymmetric theory,  they have to remain equal to the gauge
coupling, if the renormalization scheme preserves 
SUSY. Explicit computations up to three-loop order within
SUSY-QCD~\cite{Harlander:2009mn} 
confirmed this requirement for \dred{} in combination with the minimal
subtraction scheme, {\it i.e.} the  \drbar{} scheme. But,
 if \dred{} is
applied to non-supersymmetric theories, like for example QCD, this
equality is not preserved under the
renormalization~\cite{Jack:1993ws,Harlander:2006rj}.   
However, even in softly broken supersymmetric theories like the Minimal
Supersymmetric extension of the Standard Model (MSSM),
one has to worry about the \eps-scalars. In such theories, they will
receive a loop-induced mass, which will also influence
the renormalization of the genuine scalar masses. In order to decouple the
\epscalar{} masses from the $\beta$-functions of the genuine scalar
masses, additional finite counterterms proportional to the  \epscalar{}
masses have 
to be added to the  renormalized scalar masses.
This new renormalization scheme, usually  known  as the
\drbar$^{\prime}$ scheme, was introduced in
Ref~\cite{Jack:1994rk} to the one-loop order and  extended  through
two-loops in Ref.~\cite{Martin:2001vx}. 
The results presented in this letter are the same in the  \drbar{} and
\drbar$^{\prime}$ 
 schemes, because we  did not  take into account dimensionful couplings. 

As is well known, the equality of the Yukawa
couplings of gauginos to matter multiplets  and the gauge couplings, or
the equality  
 of the quartic scalar couplings, {\it e.g.} four-squark or four-slepton
couplings, and the gauge couplings  are not preserved under 
renormalization if \dreg{} is employed. This  is a direct
manifestation of the fact that \dreg{} breaks SUSY. It means that, if
one demands that the renormalized couplings are the same at some
renormalization scale, then they are different at another scale.
 This
point becomes important if we want to relate a given theory at one scale
to the same theory at another scale.
This procedure is often
  known as the running analysis and it amounts to
determine  the fundamental parameters of the MSSM solving the system of
the Renormalization Group Equations (RGEs) with two types of boundary
conditions: i) universality
conditions imposed at some very high energy scale like the unification
scale; and ii)   low-energy   constraints obtained from experiment. The
appropriate renormalization scheme at each step 
of the running analysis is not fixed {\it a priori}. In general 
the SM parameters and cross sections are mostly given in the \msbar{}
scheme~\cite{Bardeen:1978yd}, while the MSSM 
ones  are usually given  in the \drbar{} scheme. Apart from the finite shifts
of the running parameters associated with the change of renormalization
scheme, also threshold corrections, which account for the non-decoupling of
heavy particles in mass independent schemes have to be implemented.
They are known at the one-loop order for the complete
MSSM~\cite{Pierce:1996zz}, and at the
two-loop orders for the SUSY-QCD~\cite{Harlander:2005wm, Bauer:2008bj}. 

Very recently, Refs.~\cite{Signer:2008va,Signer:2005iu } have shown that
the QCD factorization theorem holds through
one-loop order,  if  \dred{} is employed in
 computations of  hadronic processes. They also provide translation
 rules  from \dred{} to other
 regularization schemes through one-loop. 
 However, it seems that the application of \dred{} to hadronic processes
 beyond  one-loop 
becomes much more involved as compared to the standard procedure based on
the DREG.
It is thus advisable to use different regularization schemes for various
parts of a practical computation. The consistency of such an approach is
guaranteed by the fact that   \dred{} and  \dreg{} are equivalent to all
orders in perturbation theory if applied to a 
renormalisable theory~\cite{dredb}. This means that the two schemes are related
by coupling constant redefinitions, under which the $\beta$-functions
calculated in one scheme transform into those computed in the other
one. In the framework of QCD, the translation rules for  the change from
\dreg{} to \dred{} 
is known up to three loops for the strong
coupling constant and the quark
masses~\cite{Harlander:2006rj,Harlander:2006xq}. In the 
MSSM, the one-loop  relations are known for the gauge, 
Yukawa,  quartic scalar couplings  and for the coupling associated with
the gaugino-chiral supermultiplet 
interactions, as well as for the gaugino masses~\cite{Martin:1993yx}. The
one-loop relation between the gauge coupling constant and the one
associated with the 
interaction of the gluino and the quark-squark multiplet has also been
verified by an on-shell computation in Ref.~\cite{Beenakker:1996dw}.  
For the strong coupling constant even the two-loop
conversion rule in SUSY-QCD is known~\cite{Harlander:2005wm}. 

It is the purpose of this letter to extend the translation ``dictionary''
between the two schemes in the framework of SUSY-CQD to two-loop order. More
precisely, we give in Section~\ref{sec::conversion} the differences
between the running 
gluino-quark-squark coupling and the running quark and gluino masses computed
in the   \msbar{} and the \drbar{} schemes at the two-loops. As a by-product
result we reconfirm the two-loop conversion   
relation derived in~\cite{Harlander:2005wm}. In
Section~\ref{sec::three} we  explicitly  
verify that the three-loop \drbar{}
$\beta$-functions and the fermion mass 
anomalous dimensions can be obtained from the \msbar{} results 
just  converting all running parameters (couplings  and
masses) according to the two-loop results derived before. In 
Appendix~\ref{sect::app} 
we discuss the one-loop renormalization of the four-squark couplings
within the \msbar{} scheme.

\section{\label{sec::conversion} Two-loop conversion rules from  \dred{}
  to \dreg{} }

In this letter, we restrict  the discussion to the translation
rules for the running parameters of
the SUSY-QCD. Thus we just
need the $SU(3)$ part of the MSSM Lagrangian. However, we give here the
results valid for a  general supersymmetric theory based on an
$SU(N)$ gauge group, with one gauge
supermultiplet in the adjoint representation $(A)$, comprising the gluon
and gluino, and  $N_f$ sets of matter 
multiplets  in the fundamental
representation $(F)$, containing the
fermions and their superpartners.\footnote{We work with Dirac fermions
  and complex scalar fields.}  

\subsection{Running coupling constants}
In order to compute the relations between  running parameters 
defined in two different renormalization schemes, one has to relate them
 to physical observables which cannot depend on the choice of 
 scheme. For example, the relationship between the strong
 coupling constant defined in the \msbar{} and \drbar{} schemes can be obtained
 from the S-matrix amplitude of a physical process involving the gauge
 coupling computed in the two schemes. 
However, beyond one-loop  the computation of the physical amplitudes
becomes very much  involved. We applied this method only for the computation of
the two-loop {\it    effective} 
 charges of the gluon-quark-quark and gluino-quark-squark couplings in
 the \drbar{} scheme, in order to prove the equality of the corresponding
 couplings at this order in perturbation theory.
 We considered the simplifying case of a
 supersymmetric theory, {\it    i.e.} massless gluino and equal-mass
 quarks and squarks and required
 the external particles to be on-shell. For the computation of the
 resulting two-loop on-shell integrals we used existing automated
 programs~\cite{onshell}.The  
   effective charges computed for on-shell gluons and gluinos are 
not infrared safe, but the infrared divergences of the two
 charges are equal. This can be understood from the fact that they are
 proportional to the corresponding one-loop {\it    effective} charges,
 which have been 
 shown to be equal~\cite{Beenakker:1996dw}, and the proportionality
 factors are universal quantities equal for gluon and gluinos in a
 supersymmetric theory.
We found that the two {\it effective}
 charges are equal, which implies that the  couplings themselves are also
 equal in the  \drbar{} scheme through two-loops. 
The equality of the
 two couplings in the \drbar{} scheme 
has been confirmed  even at the
 three-loop order in Ref.~\cite{Harlander:2009mn}. This result 
proves on the one hand the supersymmetric character of the \drbar{}
scheme,  and on the other hand it allows us to
derive the relation between the two couplings valid in the \msbar{}
scheme, as we discuss below.  \\
For the computation of the translation relations between
 the \msbar{} and \drbar{} schemes we employed  a simpler computation
 method~\cite{Harlander:2006rj}.
Starting from the observation that the
 ratio of the charge renormalization constants calculated using \dreg{}
 or \dred{}  is momentum and mass independent, one can
derive them avoiding  the use of the on-shell kinematics. Instead, one
introduces  {\it physical} 
 renormalization constants, which are computed choosing  a convenient
 kinematics for which the ``large-momentum'' or the ``hard-mass'' procedures can be
 applied,  and retains  the divergent as well as the finite pieces of
 the  renormalization constants. Up to three loops this procedure is quite
 well established ( for a detail description of the method see
 Ref.~\cite{Harlander:2006rj})  and automated programs exist to perform
 such calculations~\cite{exp,Larin:1991fz,Steinhauser:2000ry}.

Considering the {\it physical} charge of the
 gluon-quark-quark coupling at two-loop order we reconfirm the result
 derived in~\cite{Harlander:2005wm}. For completeness we reproduce it
 here
\begin{eqnarray}
\asMSbar &=& \asDRbar \bigg[ 1 - \apiDR \frac{\cA}{3} +
  \left(\apiDR\right)^2 \left(-\frac{11}{9} \cA^2  + 2 T_F N_f
    \cR\right)\bigg]\,,
\label{eq::asms}
\end{eqnarray}
where $\asMSbar=(\gsMSbar)^2/(4\pi)$ and $\asDRbar=(\gsDRbar)^2/(4\pi)$
denote the strong coupling constant in the \msbar{} and \drbar{} scheme,
respectively. 
We choose the usual normalization for the
Dynkin index $T_F$ of the fundamental representation
$Tr(T^aT^b)=T_F\delta^{ab}=\frac{1}{2}\delta^{ab}$. Accordingly, the
quadratic Casimir invariant for the fundamental representation is given
by  $C_F=T_F N_A/d(F)$, where $N_A=N^2-1$ is the number of generators
and $d(F)=N$ is the dimension of the   fundamental representation. The
Casimir invariant for the adjoint representation reads $C_A=N$.

Similarly, one can determine the conversion rules 
 for the coupling constant $\hat{\alpha}_s=(\hat{g}_s)^2/(4\pi)$ of the
 Yukawa interaction of the 
gluino and the quark-squark multiplet
\begin{eqnarray}
{\cal L}_{\tilde{g}q\tilde{q}}=-\sqrt{2}\hat{g}_s
T^a_{ij}\bigg[\bar{q}_{L,i}  \tilde{g}^a\tilde{q}_{L,j}-\bar{q}_{R,i}
   \tilde{g}^a\tilde{q}_{R,j} +h.c.\bigg]\,.
\end{eqnarray}
Here $\tilde{g},q$ and $\tilde{q}$ denote as usual the gluino, quark and
squark fields, $L$ and $R$ subscripts stand for the left- and 
right-handed components of the quark and squark fields,  and  $a$ and
$i,j$ are color 
indices of the adjoint and 
fundamental representations, respectively.\\ 
Let us remark that we performed the calculation for a general covariant
gauge and used the  cancellation of the gauge parameter in the final
results as an internal check. For the derivation of the two-loop formulae given
above, also the one-loop relation between the gauge parameter defined in the
\drbar{} and \msbar{} schemes is necessary. In order to properly take into account the Majorana character of the
gluino, the rules given in~\cite{Denner:1992vza} are applied with the
help of a  specially written PERL program~\cite{majorana}.\\

So, for the two-loop conversion rule of the gluino-quark-squark
coupling, we obtain 
\begin{eqnarray}
\ayMSbar &=& \asDRbar\bigg [ 1 + \apiDR (\cA - \cR) 
\nonumber\\
&+& \left(\apiDR\right)^2 \bigg(\frac{23}{6}
  \cA^2  - \frac{137}{12} \cA \cR
 + \frac{25}{4} \cR^2 
 + 2 T_F N_f( \cR- \cA) \bigg)\bigg]\,,
\label{eq::ayms}
\end{eqnarray}
and together with Eq.~(\ref{eq::asms}) we get the relationship between
$\ayMSbar$ and $\asMSbar$ 
\begin{eqnarray}
\ayMSbar &=& \asMSbar\bigg[ + \apiMS\left(\frac{4}{3}\cA - \cR\right) 
\nonumber\\
&+& \left(\apiMS\right)^2\left( \frac{107}{18}\cA^2  -
  \frac{145}{12}\cA\cR + \frac{25}{4}\cR^2 - 2 T_F N_f \cA  
\right)\bigg]. 
\end{eqnarray}
As a consistency check, we will show in
Section~{\ref{sec::three}} that the three-loop $\beta$-functions of
$\alpha_s$ and $\hat{\alpha_s}$ computed in the \msbar{} 
scheme can
be converted into the \drbar{} $\beta$-function~\cite{Jack:1996vg,
  Harlander:2009mn} only by means of  the finite
shifts of the running couplings.

\subsection{Running fermion masses}
 The  particle masses are other fundamental parameters of the MSSM, that
 acquired a lot of attention both theoretically and phenomenologically.  In
this letter, we provide the two-loop translation relations for the fermion
masses. They are 
functions only of the coupling constants and colour factors.
 The relations between the
running masses defined in \msbar{} and \drbar{} can be obtained using the
same requirement as for the coupling constants, that physical
observables have to be renormalization scheme independent.\\
In practice,  we have employed the easier
method of {\it physical} renormalization schemes as discussed above.
So, the running quark mass defined in the \msbar{} scheme can be
 translated into the  running  mass in the \drbar{} scheme through
\begin{eqnarray}
\mqMSbar &=& \mqDRbar\bigg[1 + \apiDR \cR +
\left(\apiDR\right)^2 \bigg(
\frac{7}{12}\cA\cR + \frac{7}{4}\cR^2 - \frac{1}{2}\cR T_F N_f
\bigg)\bigg]\,.
\label{eq::mqms}
\end{eqnarray}
For  the running gluino mass we get the following  conversion relation
\begin{eqnarray}
\mgMSbar &=& \mgDRbar\bigg[1 + \apiDR \cA +
\left(\apiDR\right)^2 \bigg(\frac{23}{6}\cA^2 - 4\cA T_F N_f +
\frac{1}{2} \cR T N_f   \bigg)\bigg]\,.
\label{eq::mgms}
\end{eqnarray}
Again, one can verify the correctness of these relations by showing that  the
three-loop mass anomalous dimensions computed in the \msbar{} scheme can be
translated into the  \drbar{} ones,  by employing only the mass and coupling
redefinitions given above. This point will be
discussed in detail in the next section.\\
 Let us point out that the relations between the running
masses defined in different renormalization schemes are free of the
renormalon problems which affects the pole masses. It is thus advisable
to  use these relations in high precision calculations of the
supersymmetric  mass spectrum.

\section{\label{sec::three} Three-loop renormalization group functions
  in DREG}
The renormalization group functions provide the scale variation of the
parameters of a quantum field theory. They have been extensively studied
and an impressive theoretical accuracy has been achieved. In the
\msbar{} scheme, the anomalous 
dimensions of all SM parameters  are known up to
two-loop level~\cite{Machacek:1983tz, Ford:1992pn}, while  for   QCD
even the four-loop order results are
available~\cite{vanRitbergen:1997va,Chetyrkin:1997dh,Vermaseren:1997fq,Czakon:2004bu}.     
 For a more general theory 
containing gauge, Yukawa and quartic scalar interactions, the gauge
$\beta$-function is known through three-loops~\cite{Pickering:2001aq}
both in the  \msbar{} and \drbar{} scheme. In the
case of the MSSM, the three-loop anomalous
dimensions for  dimensionless as well as dimensionful couplings were
derived in the \drbar{} scheme in
Refs.~\cite{Jack:1996qq,Jack:1996vg,Ferreira:1996ug}. The three-loop
anomalous dimensions for the dimensionless couplings of SUSY-QCD were
re-confirmed in Ref.~\cite{Harlander:2009mn}.

In this section, we  discuss the results for the three-loop
$\beta$-function of the gauge and
gluino-quark-squark couplings and the three-loop mass anomalous
dimensions of the quark and gluino masses in the framework of SUSY-QCD with
\msbar{} as renormalization scheme. For
such a calculation one can exploit that the divergent
part of a logarithmically divergent integral is independent of the
masses and external momenta. Thus the latter can be chosen in
a convenient way: we set to zero all masses and one of the external
momenta in the three-point functions paying attention to  not introduce
spurious infrared divergences. The resulting three-loop integrals can be
evaluated with the help of existing programs ~\cite{exp,Larin:1991fz}.
At the three-loop order in perturbation theory, the use of
$\gamma_5$ requires special care. We  adopted here the  prescription 
introduced in Ref~\cite{Harlander:2009mn}.

Apart from the technical difficulties, related to the genuine
three-loop calculation, one has to bare in mind that the couplings of
the gluino-quark-squark and four-squark interactions are different from
the gauge coupling even at the one-loop order, if the \msbar{} scheme is
employed. Since the four-squark 
couplings occur in the two-loop $\beta$-function of 
$\ayMSbar$, one needs  their one-loop renormalization constants for the
derivation of the  three-loop $\beta$-function of
$\ayMSbar$. In addition, for the conversion of this result into the
\drbar{} scheme
  the one-loop translation rules from
 \msbar{} to \drbar{}  of the four-squark couplings are
 needed. They have been known for quite some time 
 for a general renormalizable 
theory with scalars, fermions, and gauge fields at one- and two-loop
order~\cite{Cheng:1973nv, Machacek:1983tz}. 
In  SUSY-QCD the tree-level four-squark interaction is given by
\begin{eqnarray}
{\cal L}_{\tilde{q}\tilde{q}\tilde{q}\tilde{q}}=-\sum_{A,B}\frac{1}{2} g_s^2
T^a_{ij}T^a_{kl} (\tilde{q}_{L,i}^{A,\ast}\tilde{q}_{L,j}^{A}
-\tilde{q}_{R,i}^{A,\ast}\tilde{q}_{R,j}^{A})
(\tilde{q}_{L,k}^{B,\ast}\tilde{q}_{L,l}^{B}
-\tilde{q}_{R,k}^{B,\ast}\tilde{q}_{R,l}^{B})\,
\end{eqnarray}
with $A,B$ flavour indices, and $a$ and $i,j,k,l$ colour indices. At the
tree-level, the four-squark couplings are equal to the gauge
coupling. After renormalization in the \msbar{} scheme, 
one has to distinguish four types of quartic scalar couplings:
i) the coupling of  squarks with the same chirality and  flavour
$g^{A}_L\,,g^{A}_R\,,\quad$ ii) the coupling of  squarks with different
chiralities 
but the same flavour $g^{A}_{LR}\,,g^{A}_{RL} \,,\quad$ iii) the coupling
of squarks with the 
same chirality but of different flavours $g_L^{AB}\,,g_R^{AB} \,,\quad$ iv)
the coupling of 
squarks with different chiralities and flavours $g_{LR}^{AB}\,, g_{RL}^{AB}$.
Another subtlety which occurs beyond tree-level  is that the group colour
factors do not factorize, so that one has to keep track of various
colour tensors in the computation of the one-loop renormalization constants.
We introduce the following  tensors for the quartic squark couplings
{\allowdisplaybreaks
\begin{align}
(S^A_L)_{ij;kl}&=\frac{(g_L^A)^2}{4\pi}(T^a_{ij}T^a_{kl}+
T^a_{il}T^a_{kj})= \frac{(g_R^A)^2}{4\pi}(T^a_{ij}T^a_{kl}+
T^a_{il}T^a_{kj})\,,
\nonumber\\
(S^A_{LR})_{ij;kl}&=-\frac{(g^A_{LR})^2}{4\pi}(T^a_{ij}T^a_{kl})=-
\frac{(g^A_{RL})^2}{4\pi}(T^a_{ij}T^a_{kl})\,,\nonumber\\ 
(S^{AB}_{L})_{ij;kl}&=\frac{(g^{AB}_{L})^2}{4\pi}(T^a_{ij}T^a_{kl})
=\frac{(g^{AB}_{R})^2}{4\pi}(T^a_{ij}T^a_{kl})\,, 
\nonumber\\
 (S^{AB}_{LR})_{ij;kl}&=-\frac{(g^{AB}_{LR})^2}{4\pi}(T^a_{ij}T^a_{kl})
=-\frac{(g^{AB}_{RL})^2}{4\pi}(T^a_{ij}T^a_{kl})\,,  
\label{eq::tensors}
\end{align}
}and the associated coupling constants
\begin{eqnarray}
\alpha_L^{A}=\frac{(g_L^A)^2}{4\pi}\,,\quad \alpha_{LR}^{A} =
\frac{(g^A_{LR})^2}{4\pi}\,,\quad
\alpha_L^{AB} =\frac{(g^{AB}_{L})^2}{4\pi}\,,\quad
\alpha_{LR}^{AB}=\frac{(g^{AB}_{LR})^2}{4\pi} \,.
\end{eqnarray}
We provide in  Appendix~\ref{sect::app}  the one-loop \msbar{}
$\beta$-function for the  
   coupling tensors retaining the complete colour structure
   dependence. The calculation in the
\drbar{} scheme is significantly simpler since  the colour tensors
factorize. The resulting $\beta$-functions 
of scalar couplings are equal to the gauge  $\beta$-function as required
by SUSY. 

 The  translation rules for the four-squark couplings can be
 obtained from the finite pieces of the charge renormalization functions
 computed in the two schemes. We did the calculation for vanishing
 external momenta and regularized the infrared divergences  giving a
common   mass to all particles~\cite{Chetyrkin:1997fm}. 
To one-loop order they read:
\begin{eqnarray}
(S^{\delta,\overline{\rm MS}}_\lambda)_{ij;kl}&=&(S^{\delta,\overline{\rm
    DR}}_\lambda)_{ij;kl}-{\apiDR}\left(\{T^a,T^b\}_{ij}\{T^a,T^b\}_{kl}+\{T^a,T^b\}_{il}\{T^a,T^b\}_{kj}\right),\,\delta=A\,,\lambda=L\,,     
 \nonumber\\   
(S^{\delta,\overline{\rm MS}}_\lambda)_{ij;kl}&=&(S^{\delta,\overline{\rm
    DR}}_\lambda)_{ij;kl}-{\apiDR}\{T^a,T^b\}_{ij}\{T^a,T^b\}_{kl}\quad
 \mbox{otherwise}\,.
\label{eq::a4sq}
\end{eqnarray}
Here $\{T^a,T^b\}$ denotes the anti-commutator of the group generators.\\
The one-loop 
translation rules from \msbar{} to \drbar{} of the quartic scalar
couplings are known for the case of identical flavour
scalars~\cite{Martin:1993yx}. These relations coincide with those of
$(S^A_L)_{ij;kl}$ couplings in SUSY-QCD.
\subsection{Three-loop $\beta$-functions in \dreg{} }
The $\beta$-functions for the gauge and the gluino-quark-squark
couplings are defined through
\begin{eqnarray}
\betaMSbar_{\alpha_s}=\mu^2\frac{\rm{d}}{\rm{d}\mu^2}\frac{\asMSbar}{\pi}
\,,\quad\mbox{and}\quad
\betaMSbar_{\hat{\alpha}_s}=\mu^2\frac{\rm{d}}{\rm{d}\mu^2}\frac{\ayMSbar}{\pi}\,.
\end{eqnarray}
Writing
\begin{eqnarray}
\betaMSbar_{\alpha}=\sum_{i=1}^3  \beta^{\overline{\rm
    MS},(i)}_{\alpha}\,, \quad \alpha=\alpha_s\,,\hat{\alpha}_s\,, 
\end{eqnarray}
where $(i)$ stands for the loop order, we find for the gauge $\beta$-function
\begin{eqnarray}
\beta^{\overline{\rm MS},(1)}_{\alpha_s}&=&\left(\apiMS\right)^2 4 (-3\cA
+ 2 N_f T_F)\,,\nonumber\\ 
\beta^{\overline{\rm MS},(2)}_{\alpha_s}&=&\left(\apiMS\right)^3 4 (-6\cA^2
+ 8\cA N_f T_F + 12 \cF 
N_f T_F) -\left(\apiMS\right)^2\frac{\ayMSbar}{4 \pi} 16 N_f
T_F(\cA+\cF)\,,\nonumber\\
\beta^{\overline{\rm MS},(3)}_{\alpha_s}&=&\left(\apiMS\right)^4 4
\bigg[-19\cA^3 + 
  2\left(12\cA^2 + 25\cA\cR - 10\cR^2\right) N_F T_F -
  4\left(\cA + 5\cR\right) T_F^2 N_F^2\bigg]\,.\nonumber\\
\label{eq::bas}
\end{eqnarray}
In the expression for $\beta^{\overline{\rm MS},(3)}_{\alpha_s}$ as well
as in all the other three-loop formulae quoted in this letter, we
identify all couplings with 
$\asMSbar$.
The inaccuracy induced in this way is of the four-loop
order, so that the simplified formulae are enough to perform consistency
checks of the two-loop translation relations given in the previous section.
In practice, we derived the formulae distinguishing between the various
couplings, but the results are too long to be presented here. 
The three-loop results with complete dependence on different couplings 
 can be obtained in electronic form from the author.\\

The three-loop $\beta$-function of the gluino-quark-squark coupling
reads
{\allowdisplaybreaks
\begin{align}
\beta^{\overline{\rm MS},(1)}_{\hat{\alpha_s}}&=
-\apiMS\frac{\ayMSbar}{4 \pi} 12 (\cA + \cR) +
   \left(\frac{\ayMSbar}{4 \pi}\right)^2 4 (3\cR + 2 T_F N_F)\,,\nonumber\\
\beta^{\overline{\rm MS},(2)}_{\hat{\alpha_s}}&=
\left(\frac{\ayMSbar}{4 \pi}\right)^3  4 \left[4\cA^2 - 12\cA\cR + 2 \cR
  (\cR - 7 N_F T_F)\right]
\nonumber\\
&+\left(\frac{\ayMSbar}{4 \pi}\right)^2 4 \bigg[
\left(\apiMS\right) \frac{6\cA^2 + 9\cA\cR + 21\cR^2 - 2\cA T_F
N_F + 34\cR T_F N_F}{2}\nonumber\\
&-4\sum_{\delta=A,AB}\frac{(T^aT^b)_{ji}(S^\delta_L)_{ij;kl}(T^bT^a)_{lk}}{N_A
  T_F}
-4\sum_{\delta=A,AB}\frac{(T^aT^b)_{ji}(S^\delta_{LR})_{ij;kl}
(T^aT^b)_{lk}}{N_A T_F}
\bigg]\nonumber\\
&
+\left(\frac{\ayMSbar}{4 \pi}\right) 4 \bigg[
\left(\apiMS\right)^2(-16\cA^2 + \frac{11}{4}\cA\cR - 6\cR^2 + 7\cA N_F
T_F + 4\cR N_F T_F)\nonumber\\
&
+\frac{\cR}{2}\sum_{\delta=A,AB}\sum_{\lambda=L,LR}
\frac{(S^\delta_\lambda)_{ij;kl}(S^\delta_\lambda)_{ji;lk}}{N_A   T_F} 
\bigg]
\nonumber\\
&=
\left(\frac{\ayMSbar}{4 \pi}\right)^3  4 \left[4\cA^2 - 12\cA\cR + 2 \cR
  (\cR - 7 N_F T_F)\right]
\nonumber\\
&+\left(\frac{\ayMSbar}{4 \pi}\right)^2 4 \bigg[
\left(\apiMS\right) \frac{6\cA^2 + 9\cA\cR + 21\cR^2 - 2\cA T_F
N_F + 34\cR T_F N_F}{2}\nonumber\\
&-\left(\ala\right) ( \cA  T_F + 16 D_3(F) T_F 
+ \cA^2 - 4\cA\cR + 4\cR^2 ) -
   \left(\alra\right) (\cA - 16 D_3(F)) T_F 
\nonumber\\
&
- \left(\alab\right) (\cA + 16 D_3(F))
   T_F N_Q - \left(\alrab\right) (\cA - 16 D_3(F) ) T_F N_Q 
\bigg]\nonumber\\
&
+\left(\frac{\ayMSbar}{4 \pi}\right) 4 \bigg[
\left(\apiMS\right)^2(-16\cA^2 + \frac{11}{4}\cA\cR - 6\cR^2 + 7\cA N_F
T_F + 4\cR N_F T_F)\nonumber\\
&  + \left(\ala\right)^2\cR\frac{2 T_F - \cA + 2\cR}{4} +
\left(\alra\right)^2\frac{\cR T_F}{2}
\nonumber\\
&
+ \left(\alab\right)^2\frac{\cR  N_Q T_F}{2}
 +\left(\alrab\right)^2\frac{\cR N_Q T_F}{2}
\bigg]\,,
\nonumber\\
    \beta^{\overline{\rm MS},(3)}_{\hat{\alpha_s}}
&=
\left(\apiMS\right)^4 4 \bigg[ 
-\frac{188}{3}\cA^3 
+ \frac{167}{3}\cA^2\cR + \frac{145}{3} \cA^2 N_F T_F
+ \frac{17}{4}\cR^2\cA 
\nonumber\\
&
+\frac{107}{2}\cR \cA N_F T_F
-8\cA N_F^2 T_F^2
- \frac{81}{4}\cR^3 
- \frac{49}{2}\cR^2 N_F T_F -20\cR N_F^2 T_F^2 \nonumber\\
&
- 32 D_4(FA)  -128  D_4(FF) N_F T_F  
\bigg]\,,
\label{eq::bay}
\end{align}
}where  $N_Q=N_F-1$ counts the number of quark/squark flavours $B$ different
 from the external quark/squark flavour $A$.
The additional colour factors occurring in the above results are defined as
    \begin{eqnarray}
 D_3(F) &=&\frac{d_F^{abc}d_F^{abc}}{N_A}=\frac{N^2-4}{16 N}\,,
\qquad D_4(FA) = \frac{d_F^{abcd}d_A^{abcd}
}{N_A}=\frac{N(N^2+6)}{48}\,,\nonumber
\\
 D_4(FF) &=& \frac{d_F^{abcd}d_F^{abcd} }{N_A}
=\frac{18-6 N^2 + N^4 }{96 N^2}\,,
    \end{eqnarray}
where $d_F^{abc}, d_F^{abcd}, d_A^{abcd}$ are the fully symmetric rank three
and four tensors of SU(N), as defined in
Ref.~\cite{vanRitbergen:1998pn}.
\\ 
$\beta^{\overline{\rm MS},(2)}_{\hat{\alpha_s}}$ is given first as a
function of the quartic scalar coupling tensors. Implementing their
explicit expressions (\ref{eq::tensors}) one gets the RHS of the second
equality sign. The explicit dependence on the coupling tensors is needed
  for the computation of the three-loop result $\beta^{\overline{\rm
      MS},(3)}_{\hat{\alpha_s}}$ as a function of the different types of
  couplings. As can be understood from the above formulae,  when the
  quartic scalar coupling tensors occurring in the two-loop diagrams 
  are renormalized,  their one-loop renormalization functions are 
  contracted with three different colour structures. We did the
  renormalization at the diagram level, employing the appropriate
  colour projectors.   For the derivation of the
  three-loop results with all couplings set to be equal to
  $\alpha_s^{\overline{\rm MS}}$, one can avoid the introduction of
  coupling tensors 
  for the four-squark interaction. In this case the colour structures of the
  tree-level  couplings are preserved under the
  renormalization to the one-loop order and so, their renormalization 
  can be done as usual. However, for the conversion of  the three-loop
  \msbar{}   results to the \drbar{} scheme the introduction of the
  coupling tensors is unavoidable, because the colour structures do not
  factorize in the second  equation of the translation relations
  (\ref{eq::a4sq}).\\ 
Furthermore, it is a straightforward calculation to show that employing the
conversion rules given 
in Eqs.~(\ref{eq::asms}), (\ref{eq::ayms}), (\ref{eq::a4sq}) into the
\msbar{} three-loop 
$\beta$-functions (\ref{eq::bas}) and  (\ref{eq::bay}), one obtains the
\drbar{}  
$\beta$-function computed in
Refs.~\cite{Jack:1996vg,Harlander:2009mn}. Since the couplings $\alpha_s$ and
$\hat{\alpha}_s$ occur already at the one-loop order, their
two-loop translation rules are necessary to convert the three-loop
$\beta$-functions from \msbar{} to \drbar{}. 
 This is a strong consistency check for the translation
rules we discussed in the previous section.

\subsection{Three-loop  fermion mass anomalous dimensions in
  \dreg{}\label{app::beta4sq} } 
In this section we provide the fermion (quark and gluino) mass anomalous
dimensions within the \msbar{} scheme through three-loops. They are
derived from the renormalization constants of the fermion
masses, which can be calculated by decomposing the fermion self-energy
into its vector and scalar parts and then computing the   
counterterms for  the wave functions and masses.
We define the fermion (quark or gluino) mass anomalous dimensions as
\begin{eqnarray}
    \gamma_A = \frac{\mu^2}{m_A}\frac{\rm d m_A}{\rm d \mu^2 }\,,\quad
    A=q,\tilde{g}\,. 
\end{eqnarray} 
Writing their  expansion in the perturbation theory like
\begin{eqnarray}
 \gamma_A = \sum_{i=1}^3 \gamma_{A}^{\overline{\rm MS},(i)}\,,
\end{eqnarray}
we have for the quark mass anomalous dimension 
{\allowdisplaybreaks
\begin{align}
 \gamma_{q}^{\overline{\rm MS},(1)} &=
  \left(\frac{\ayMSbar}{4
   \pi}\right) \cR -\left(\apiMS\right) 3  \cR\,,\nonumber\\
\gamma_{q}^{\overline{\rm MS},(2)} &=
\left(\apiMS\right)^2(-\frac{29}{2}\cA\cR - \frac{3}{2}\cR^2 + 7\cR N_F
T_F) 
\nonumber\\
&
+ \left(\frac{\ayMSbar}{4\pi}\right)^2(-2\cA\cR + \cR^2 - \cR N_F T_F)
+\left(\apiMS\right)\left(\frac{\ayMSbar}{4\pi}\right)\frac{11}{2}
(\cA\cR +\cR^2)\,,\nonumber\\
\gamma_{q}^{\overline{\rm MS},(3)} &=
\left(\apiMS\right)^3\bigg[
-\frac{115}{3}\cA^2\cR + \frac{43}{4}\cA\cR^2 - \frac{59}{4}\cR^3 +
6\cR N_F^2T_F^2
\nonumber\\
&
  +  (14\cA\cR + 47\cR^2 + 48\cA\cR \zeta(3) - 48\cR^2\zeta(3)) N_F T_F
\bigg]\,,
\end{align}
}where $\zeta$ denotes the Riemann's zeta function with
$\zeta(3)=1.20206$.
For the gluino mass anomalous dimension we obtain
{\allowdisplaybreaks
\begin{align}
\gamma_{\tilde{g}}^{\overline{\rm MS},(1)} &=
\left(\apiMS\right)
(-3)\cA + \left(\frac{\ayMSbar}{4\pi}\right) 2 N_F T_F\,,\nonumber\\
\gamma_{\tilde{g}}^{\overline{\rm MS},(2)} &=
 \left(\apiMS\right)^2
\cA (-16\cA +7 N_F T_F) 
-\left(\frac{\ayMSbar}{4\pi}\right)^2(4\cA - \cR) N_F T_F 
\nonumber\\
&
+ \left(\apiMS\right) \left(\frac{\ayMSbar}{4\pi}\right) (5\cA + 17\cR)
N_F T_F \,, 
\nonumber\\
\gamma_{\tilde{g}}^{\overline{\rm MS},(3)} &=
 \left(\apiMS\right)^3
\bigg[-\frac{310}{3}\cA^3 + (103\cA^2 
 + \frac{347}{2}\cA\cR - \frac{83}{2}\cR^2) N_F T_F 
\nonumber\\
&
+(-24\cA - 74\cR) N_F^2 T_F^2
   \bigg]\,.
\end{align}
}It is an easy exercise to verify that the three-loop \msbar{} mass
anomalous 
dimensions given above differ from the ones computed in \drbar{} 
scheme~\cite{Jack:1996qq,Harlander:2009mn} only by the finite shifts for
coupling constants and masses discussed in
Section~\ref{sec::conversion}. Let us point out that, for the conversion
of the three-loop mass anomalous dimensions the two-loop relations for
masses and couplings are needed. So, this provide us  with another important
consistency 
check of Eqs.~(\ref{eq::asms}, \ref{eq::ayms}, \ref{eq::mqms}, \ref{eq::mgms}).

\section{\label{sec::concl}Conclusions}
In this letter we present the two-loop translation rules between
\drbar{} and \msbar{} scheme  for the running  gluino-quark-squark coupling and
for the gluino and quark masses. We also confirm the two-loop relation
for the gauge coupling given in
Ref.~\cite{Harlander:2005wm}. Furthermore, we prove that the three-loop 
$\beta$-function of the gauge and gluino-quark-squark couplings and the
anomalous dimensions of the quark and gluino masses calculated in the
\msbar{} scheme can be converted into the known  \drbar{} results, 
by means of  these two-loop parameter redefinitions. This is a powerful 
consistency check of our two-loop results.\\
As a by-product of our calculation, we derive the one-loop RGEs for the
four-squark coupling in the \msbar{} scheme and their conversion rules
to the \drbar{} scheme.   
\begin{appendix}

\section{\label{sect::app} One-loop $\beta$-functions of the four-squark
  couplings in   \msbar{}}
As mentioned before, the four-squark couplings behave as tensors in
colour space. Their one-loop $\beta$-functions in the \msbar{} scheme read
{\allowdisplaybreaks
\begin{align}
\mu^2\frac{\rm{d}}{\rm{d}\mu^2} \frac{(S^{\delta,\overline{\rm MS}}_\lambda)_{ij;kl}}{\pi}&=
\frac{1}{4\pi^2}(\Lambda^\delta_\lambda)_{ij;kl} - 32\left(\frac{\ayMSbar}{4
  \pi}\right)^2 (H^\delta_\lambda)_{ij;kl}+ 
4\left(\frac{\ayMSbar}{4 \pi}\right) \frac{(S^\delta_\lambda)_{ij;kl}}{\pi} \cR
\nonumber\\
&  +
6
\left(\apiMS\right)^2 (G^\delta_\lambda)_{ij;kl} 
-6 \left(\apiMS\right) \frac{(S^\delta_\lambda)_{ij;kl}}{\pi}\cR
+\frac{1}{4\pi^2}
(\Omega^\delta_\lambda)_{ij;kl}\,,
\nonumber\\
&
\quad
\mbox{with}\quad \delta=A,AB\,,\quad \mbox{and} \quad \lambda=L,LR\,.
\label{eq::betaa4sq}
\end{align}
}
The new colour tensors are defined as follows
{\allowdisplaybreaks
\begin{align}
(\Lambda^A_L)_{ij;kl}&=
(S^A_L)_{ij;mn}(S^A_L)_{nm;kl} +
(S^A_L)_{il;mn}(S^A_L)_{nm;kj}+\frac{1}{2}(S^A_L)_{im;kn}(S^A_L)_{mj;nl}
\,,\nonumber\\ 
(H^A_L)_{ij;kl} &= 
(T^a T^b)_{ij} (T^b T^a)_{kl}+ (T^a T^b)_{kj} (T^b  T^a)_{il}\,,
\nonumber\\
(G^A_L)_{ij;kl} &=
\{T^a,T^b\}_{ij} \{T^a,T^b\}_{kl}+\{T^a,T^b\}_{il}\{T^a,T^b\}_{kj}\,
\nonumber\\
(\Omega^A_L)_{ij;kl} &=
(S^A_{LR})_{ij;mn}(S^A_{LR})_{nm;kl}+(S^A_{LR})_{il;mn}(S^A_{LR})_{nm;kj}
\nonumber\\
&+
\sum_{B\ne A}\bigg[(S^{AB}_{L})_{ij;mn}(S^{AB}_{L})_{nm;kl}
+(S^{AB}_{L})_{il;mn}(S^{AB}_{L})_{nm;kj}\bigg]
\nonumber\\
&+
\sum_{B\ne A}\bigg[(S^{AB}_{LR})_{ij;mn}(S^{AB}_{LR})_{nm;kl}
+(S^{AB}_{LR})_{il;mn}(S^{AB}_{LR})_{nm;kj}\bigg]\,,\nonumber\\
\nonumber\\
(\Lambda^A_{LR})_{ij;kl}&=
 (S^A_{LR})_{im;nl}(S^A_{LR})_{mj;kn} +
 (S^A_{LR})_{im;kn}(S^A_{LR})_{mj;nl}\,,
\nonumber\\
(H^A_{LR})_{ij;kl} & =
(T^a T^b)_{ij} (T^a T^b)_{kl}\,,
\nonumber\\
(G^A_{LR})_{ij;kl} & =
\{T^a,T^b\}_{ij} \{T^a,T^b\}_{kl}\,,
\nonumber\\
(\Omega^A_{LR})_{ij;kl} &=
2 (S^A_{L})_{ij;mn}(S^A_{LR})_{nm;kl}+
2 \sum_{B \ne A}\,(S^{AB}_{L})_{ij;mn}(S^{AB}_{LR})_{nm;kl}\,,\nonumber\\
\nonumber\\
(\Lambda^{AB}_{L})_{ij;kl}&=
(S^{AB}_{L})_{im;nl}(S^{AB}_{L})_{mj;kn} +
(S^{AB}_{L})_{im;kn}(S^{AB}_{L})_{mj;nl}\,,
\nonumber\\
(H^{AB}_{L})_{ij;kl} & =
(T^a T^b)_{ij} (T^b T^a)_{kl}\,,
\nonumber\\
(G^{AB}_{L})_{ij;kl} & =G^A_{LR})_{ij;kl}\,,
\nonumber\\
(\Omega^{AB}_{L})_{ij;kl} &=
\sum_{C} (S^{AC}_{L})_{ij;mn}(S^{BC}_{L})_{nm;kl}+
\sum_{C} (S^{AC}_{LR})_{ij;mn}(S^{BC}_{LR})_{nm;kl}\,,\nonumber\\
\nonumber\\
(\Lambda^{AB}_{LR})_{ij;kl}&=
(S^{AB}_{LR})_{im;nl}(S^{AB}_{LR})_{mj;kn} +
(S^{AB}_{LR})_{im;kn}(S^{AB}_{LR})_{mj;nl}\,,
\nonumber\\
(H^{AB}_{LR})_{ij;kl} & =(H^{A}_{LR})_{ij;kl} \,,
\nonumber\\
(G^{AB}_{LR})_{ij;kl} &=(G^A_{LR})_{ij;kl}\,,
\nonumber\\
(\Omega^{AB}_{LR})_{ij;kl} &=
\sum_{C} (S^{AC}_{L})_{ij;mn}(S^{BC}_{LR})_{nm;kl}+
\sum_{C}\,(S^{AC}_{LR})_{ij;mn}(S^{BC}_{L})_{nm;kl}\,.
\end{align}
}

As can be easily verified, even if we identify the four types of quartic scalar
interactions their one-loop $\beta$-functions remain different.
If in addition, one sets them equal to the gauge coupling and to the
gluino-squark-quark coupling equal, {\it i.e.} if the 
\drbar{} scheme constraints are fulfilled, then the colour structures
factorize. The resulting one-loop 
$\beta$-functions for the scalar couplings  are identical with the one-loop
\drbar{} gauge $\beta$-function, as required by SUSY.       
\end{appendix}
\vspace*{1em}

\noindent
{\large\bf Acknowledgements}\\ 
I would like to thank R.~Harlander and M.~Steinhauser for reading the
manuscript and for valuable discussions and useful comments.\\
 This work was supported by the DFG through SFB/TR~9.

\end{document}